\documentclass[%
 reprint,
preprintnumbers,
 amsmath,amssymb,
 aps, groupedaddress, superscriptaddress,
]{revtex4-2}

\pdfoutput=1

\usepackage[dvipsnames]{xcolor}
\usepackage{graphicx}
\usepackage{dcolumn}
\usepackage{bm}
\usepackage{hyperref}
\usepackage{multirow}
\usepackage{soul}
\usepackage{color}
\usepackage{lineno}
\usepackage[normalem]{ulem}

\definecolor{myblue}{rgb}{0.152941176,0.549019608,0.670588235}
\definecolor{newred}{cmyk}{0,1,1,0.2}
\definecolor{newblue}{cmyk}{1,1,0,0.1}

\definecolor{linkcolor}{rgb}{0.6,0,0}
\definecolor{citecolor}{rgb}{0,0,0.75}
\definecolor{urlcolor}{rgb}{0.12,0.46,0.7}

\hypersetup{   
    pdftoolbar=true,        
    pdfmenubar=true,        
    pdffitwindow=false,
    pdfnewwindow=true,
    urlcolor=urlcolor, 
    linkcolor=linkcolor,
    citecolor=citecolor,
    colorlinks=true,
    linktocpage=true
}

\def\equationautorefname~#1\null{Eq.\,(#1)\null}
\newcommand{\appendixref}[1]{\hyperref[#1]{appendix~\ref{#1}}}

\usepackage{enumitem}
\usepackage{braket}
\usepackage{hyperref}
\usepackage[bottom]{footmisc}
\usepackage{amsmath,amssymb,amsfonts,mathrsfs}
\usepackage{booktabs}
\usepackage{mathtools}
\usepackage{bm}
\usepackage{dcolumn}
\usepackage{graphicx}   
\usepackage[latin1]{inputenc}
\usepackage{latexsym}
\usepackage{color}
\usepackage[acronym]{glossaries}

\usepackage[caption=false]{subfig}

\newcommand{\hecho}{h^\T{echo}}

\newcommand{\Hp}{\mathscr H^+}

\newcommand{\RBH}{\mathcal{R}^{\text{BH}}}
\newcommand{\RQBH}{\mathcal{R}^\T{QBH}}

\newcommand{\Zin}{Z^{\text{in}}_{\ell m \omega}}
\newcommand{\Zouteco}{Z^{\text{out QBH}}_{\ell m \omega}}

\newcommand{\Zout}{Z^{\text{out}}_{\ell m \omega}}
\newcommand{\Din}{D^{\text{in}}_{\ell m \omega}}
\newcommand{\Dout}{D^{\text{out}}_{\ell m \omega}}
\newcommand{\Cin}{C^{\text{in}}_{\ell m \omega}}
\newcommand{\Cout}{C^{\text{out}}_{\ell m \omega}}
\newcommand{\Zinf}{Z^{\infty}_{\ell m \omega}}
\newcommand{\Yin}{Y^{\text{in}}_{\ell m \omega}}
\newcommand{\Yineco}{Y^{\text{in QBH}}_{\ell m \omega}}
\newcommand{\Yout}{Y^{\text{out}}_{\ell m \omega}}
\newcommand{\Yinf}{Y^{\infty}_{\ell m \omega}}
\newcommand{\T}[1]{\text{#1}}

\newcommand{\scrip}{{\mathscr I}^+}
\newcommand{\scrim}{{\mathscr I}^-}
\newcommand{\RNum}[1]{\uppercase\expandafter{\romannumeral #1\relax}}
\newcommand{\Msol}{\ensuremath{M_{\odot}}}

\newcommand{\CornellPhysics}{\affiliation{Department of Physics, Cornell University, Ithaca, NY, 14853, USA}}
\newcommand{\Cornell}{\affiliation{Cornell Center for Astrophysics and Planetary Science, Cornell University, Ithaca, New York 14853, USA}}
\newcommand{\CornellLepp}{\affiliation{Laboratory for Elementary Particle Physics, Cornell University, Ithaca, New York 14853, USA}}
\newcommand{\Caltech}{\affiliation{Theoretical Astrophysics 350-17, California Institute of Technology, Pasadena, CA 91125, USA}}

\makeglossaries
    
\newacronym{e2e}{E2E}{End-To-End}
\newacronym{inrep}{INREP}{Initial Noise REduction Pipeline}
\newacronym{tdi}{TDI}{Time Delay Interferometry}
\newacronym{ttl}{TTL}{Tilt-To-Length couplings}
\newacronym{dfacs}{DFACS}{Drag-Free and Attitude Control System}
\newacronym{ldc}{LDC}{LISA Data Challenge}
\newacronym{lisa}{LISA}{the Laser Interferometer Space Antenna}
\newacronym{emri}{EMRI}{Extreme Mass Ratio Inspiral}
\newacronym{ifo}{IFO}{Interferometry System}
\newacronym{grs}{GRS}{Gravitational Reference Sensor}
\newacronym{tmdws}{TM-DWS}{Test-Mass Differential Wavefront Sensing}
\newacronym{ldws}{LDWS}{Long-arm Differential Wavefront Sensing}
\newacronym[	plural={MOSAs},
		        first={Moving Optical Sub-Assembly},
		        firstplural={Moving Optical Sub-Assemblies}
            ]{mosa}{MOSA}{Moving Optical Sub-Assembly}
\newacronym{siso}{SISO}{Single-Input Single-Output}
\newacronym{mimo}{MIMO}{Multiple-Input Multiple-Output}
\newacronym[plural=MBHB's, firstplural=Massive Black Holes Binaries (MBHB's)]{mbhb}{MBHB}{Massive Black Holes Binary}
\newacronym{cmb}{CMB}{Cosmic Microwave Background}
\newacronym{sgwb}{SGWB}{Stochastic Gravitational Waves Background}
\newacronym{pta}{PTA}{Pulsar Timing Arrays}
\newacronym{gw}{GW}{Gravitational Wave}
\newacronym{snr}{SNR}{Signal-to-Noise Ratio}
\newacronym{pbh}{PBH}{Primordial Black Holes}
\newcommand{\orcid}[1]{\href{https://orcid.org/#1}{\includegraphics[width=8pt]{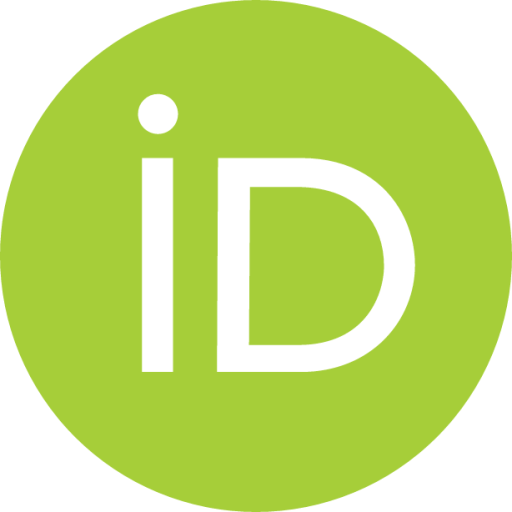}}}

\begin{document}

\title{Echoes from beyond: detecting gravitational-wave quantum imprints with LISA}

\author{Nils Deppe \orcid{0000-0003-4557-4115}} \CornellLepp \CornellPhysics \Cornell

\author{Lavinia Heisenberg}
\affiliation{Institute for Theoretical Physics, University of Heidelberg, Philosophenweg 16,
69120 Heidelberg,
Germany} 

\author{Henri Inchausp\'e \orcid{0000-0002-4664-6451}}
\affiliation{Institute for Theoretical Physics, KU Leuven,
Celestijnenlaan 200D, B-3001 Leuven, Belgium}
\affiliation{Leuven Gravity Institute, KU Leuven,
Celestijnenlaan 200D box 2415, 3001 Leuven, Belgium}

\author{Lawrence E.~Kidder \orcid{0000-0001-5392-7342}} \Cornell

\author{David Maibach \orcid{0000-0002-5294-464X}}
\email{d.maibach@thphys.uni-heidelberg.de}
\affiliation{Institute for Theoretical Physics, University of Heidelberg, Philosophenweg 16,
69120 Heidelberg,
Germany} 

\author{Sizheng Ma \orcid{0000-0002-4645-453X}}
\affiliation{Perimeter Institute for Theoretical Physics, Waterloo, ON N2L2Y5, Canada} 

\author{Jordan Moxon \orcid{0000-0001-9891-8677}} \Caltech
\author{Kyle C.~Nelli \orcid{0000-0003-2426-8768}} \Caltech
\author{William Throwe \orcid{0000-0001-5059-4378}} \Cornell
\author{Nils L.~Vu \orcid{0000-0002-5767-3949}} \Caltech


\begin{abstract}

We revisit gravitational wave echoes as a pathway to measuring quantization effects of black hole horizons. In particular, we construct a robust and phenomenologically motivated toy model for black hole reflectivity, enabling us to estimate the detectability of waveform echoes in gravitational wave interferometer data. We demonstrate how the frequency information of echoes, upon detection, can be leveraged to probe and constrain the quantum scale associated with the underlying theory of Quantum Gravity. Our analysis shows that echoes may manifest in LISA data with an estimated signal-to-noise ratio of up to $\mathcal{O}(10^2)$. For such signals, the frequency features indicative of horizon quantization can be determined with sub-percent precision.

\end{abstract}

\maketitle

\section{Introduction}

Classically, the black hole (BH) horizon is considered an unremarkable region of spacetime, with no distinctive effects expected locally. However, in the quantum regime, this picture may change dramatically, as indicated initially by Hawking's discovery of thermal radiation \cite{Hawking:1975vcx}. Although the precise quantum structure of a BH's horizon remains unknown, numerous theories have been proposed to model its quantum properties. These include the membrane paradigm \cite{Damour_book, Throne_book, PhysRevD.33.915} (see also \cite{Maggio_2020, Chakraborty_2022, Rahman_2021}), effective field theories \cite{Burgess_2018} and phenomenological approaches \cite{Oshita_2020} (see \cite{PhysRevD.105.044046} for quantum mechanics inspired modeling techniques).  
Another compelling ansatz based on the Bekenstein-Mukhanov model of BH quantization has been recently proposed in \cite{Agullo_2021}: Given an area discretization $A_N = \alpha \ell_\T{Pl}^2 N$, where $\ell_\T{Pl}$ is the Planck length, $N$ a positive integer and $\alpha$ denotes a phenomenological constant determined by the underlying theory of Quantum Gravity (QG) ($\alpha\in \mathbb R$\footnote{In Bekenstein's original works, a value of $\alpha=8\pi$ is adapted. Interestingly, the same value is recovered by various unrelated derivations (see for instance \cite{Maggiore_2008}).}), a discrete mass spectrum is implied. This in turn leads to the quantization of BH emission and absorption processes. Concretely, only frequencies matching the mass gaps
\begin{align}
\label{equ:char_freq_init}
    \omega = \frac{|\Delta M|}{\hbar} = \frac{\alpha \Delta N}{32\pi M}\,,
\end{align}
can be absorbed or emitted. $M$ thereby refers to the mass of the BH and $\Delta N$ to the number of microstates. As these frequencies scale in $1/M$, Planck-scale effects are magnified, pushing them into the realm of detectability for gravitational wave (GW) interferometers \cite{Agullo_2021}. For spinning macroscopic BHs, i.e., $N\gg1$, one can compute the characteristic frequencies of the BH as a function of the phenomenological constant, $\alpha$, and dimensionless spin, $a$, as \cite{Agullo_2021}
\begin{align}
\label{equ:characteristic_frequency}
    \omega_N(\alpha,a)= \frac{\kappa \alpha N}{8\pi} + 2\Omega_H + \mathcal{O}(N^{-1})\,,
\end{align}
where $\kappa = \sqrt{1-a^2}/[2M(1+\sqrt{1-a^2})]$ and $\Omega_H = a/[2M(1+\sqrt{1-a^2})]$. Here, $\kappa$ and $\Omega_H$ denote the surface gravity and the angular momentum, respectively.

The discrete spectrum implies that a significant portion of radiation, including GWs, may not be absorbed by a quantum black hole (QBH). As discussed in \cite{Cardoso_2019, Foit_2019}, the unabsorbed modes could be reflected, resulting in a late-time echo in gravitational waveforms \cite{Cardoso_2016,Cardoso_2016_II,Cardoso_2017} (see also \cite{Afshordi_I,Afshordi_II,Damico_2020,Manikandan_2022, Chakraborty_2022}; we point out \cite{fransen2024gravitationalwavesignaturesdepartures} for a more generalized treatment of quantum modifications such as echos and \cite{kokkotas1996pulsatingrelativisticstars, Tominaga_1999, Ferrari_2000,Seb_Voe_I,Seb_Voe_II} for earlier works on echoes). 
A delayed GW echo reaching the detector after radiation from a (classically) merging binary BH (BBH) has passed through is generally predicted by a manifold of phenomenologies. This includes deviations from General Relativity (GR) \cite{Zhang_2018,Dong_2021}, the presence of near-horizon (quantum) structure \cite{Almheiri_2013, Giddings_2016, Oshita_2019, Abedi_2023,Chakravarti_2021,Cardoso_2019,Wang_2020, Oshita_2020} as well as the existence of exotic compact objects (ECOs) replacing the classical BHs \cite{Mazur_2004,Visser_2004,Damour_2007,Mathur_2005, Holdom_2017}. In particular, in the context of near-horizon quantum structure, the echo manifests irrefutable evidence of an interplay between quantum physics and gravity \cite{Chakraborty_2022, Oshita_2020, Oshita_2019}. While the existence of echoes is generally still under debate from a conceptual point of view \cite{Chakravarti:2023wlc,zimmerman2023rogueechoesexoticcompact} (also \cite{Maggio_2019, shit_1, shit_2, shit_3, shit_4, shit_5, chen2019instabilityexoticcompactobjects}), the potential for detection of these signatures by current and future GW interferometers has been addressed in the literature, e.g., \cite{Laghi:2020rgl,Arun_2022,Uchikata:2023zcu, Testa_2018, Maggio_2019_II} (see also \cite{Abedi_2019,abedi2020echoesabyssstatusupdate}). However, these discussions mostly establish detection forecasts and do not expound upon physical information hidden in the echo's strain signal. In this work, we present a novel case study on the detection of echo-like signatures arising due to area quantization arguments with \gls{lisa} \cite{amaro-seoane_laser_2017} applying realistic data analysis techniques and providing a rigorous assessment of quantization features within the signal. To our knowledge, this work constitutes an unprecedented search for features of fundamental quanta of the underlying theory of QG in GW data, demonstrating that GW echoes could serve as a smoking gun for advancing our understanding of BH physics.

Throughout this work, we adapt the following assumptions: i) GR is the effective theory describing the propagation of GWs throughout spacetime, ii) in the transition from classical to QBH, the ``quantumness'' manifest by discretizing the mass spectrum of BHs, iii) the radiation directed towards the BH is sourced during the ringdown phase of a BBH merger, iv) of such radiation, the unabsorbed portion is reflected off the BH.

\section{Reflectivity model}
To accommodate iv), we adopt a simple yet robust phenomenological reflectivity model that captures a wide range of echo morphologies while remaining agnostic regarding the underlying quantization scale. Our model is based on the assumption that the characteristic frequencies $\omega_N(\alpha, a)$ appear as sharp lines in the QBH reflectivity spectrum resembling an atom-like description for the BH as a direct consequence of the Bekenstein-Mukhanov proposal \cite{Bekenstein1997QuantumBH} (see also \cite{Cardoso_2019,Foit_2019} for further motivation). Treating the BH as an excited multilevel quantum system, spectral line broadening may occur for various reasons. As demonstrated in \cite{Agullo_2021}, the absorption lines' thickness increases, for instance, with the QBH's spin. Moreover, the presence of quantum fluctuations and uncertainty close to the horizon can turn the sharp roots of the reflectivity coefficient into wider cusps around frequencies $\omega_N$. Examples of such quantum effects are discussed in the context of quantum-related modifications of the quasi-normal mode (QNM) spectrum \cite{Berti_2009, Konoplya_2011} and Hawking radiation \cite{Amelino_Camelia_2001, York_1986, Giddings_2016}.

Based on these assumptions, a first ansatz for the reflectivity reads
\begin{align}
    \RQBH \sim \begin{cases} 
            1, & \omega\leq\omega_1 \,, \\
             \left|\sin{\left( \frac{\pi\omega}{\omega_N(\alpha)} \right)}\right|^\delta\,, & \omega > \omega_1 \,,
            \end{cases}
\end{align}
where $\delta$ parameterizes the ``sharpness'' of the spectral lines associated to $\omega_N$. Assuming an equal spacing between the characteristic frequencies allows for the use of a periodic function. As the analysis below focuses primarily on the identification of first characteristic frequency $\omega_1$, the exact spacing does not affect our considerations besides the concrete functional description of $\RQBH$. 

On a quantum level, the reflection of ingoing GWs (with respect to the horizon) is not the only emission channel for a BH. Already on a semi-classical level, Hawking radiation enters, establishing an emission channel that is equally susceptible to the quantization arguments \cite{Bekenstein_1995, Hod_2015}. In this context, the frequency corresponding to the Hawking temperature $T_H$ commonly represents a bound beyond which the reflectivity is exponentially suppressed \cite{Abedi_2017,Oshita_2019, Oshita_2020, Chen_2021,  Chakraborty_2022}. This assumption is frequently adapted in the realm of ECOs and, in particular, is in concordance with the concept of BH microstates \cite{Mathur_2014, Chowdhury_2013}. Additional support for a suppression factor comes from astrophysical observations \cite{Broderick_2009, Broderick_2015} and discussions involving solutions to the BH information paradox \cite{Afshordi_2016, Pen_2014}. Thus, we adopt a conservative exponential suppression factor of $\exp{[-|\omega|/(2T_{QH})]}$ in our ansatz for the reflectivity. Staying agnostic with respect to an alteration of the cut-off temperature, we replace the Hawking temperature $T_H$ with $T_{QH} = \epsilon T_H$, where $T_H= 1/8\pi$. For $T_{QH} \rightarrow 0$, the reflectivity is fully suppressed and the classical limit is recovered.

\begin{figure}
	\centering
	\includegraphics[width=0.8\linewidth]{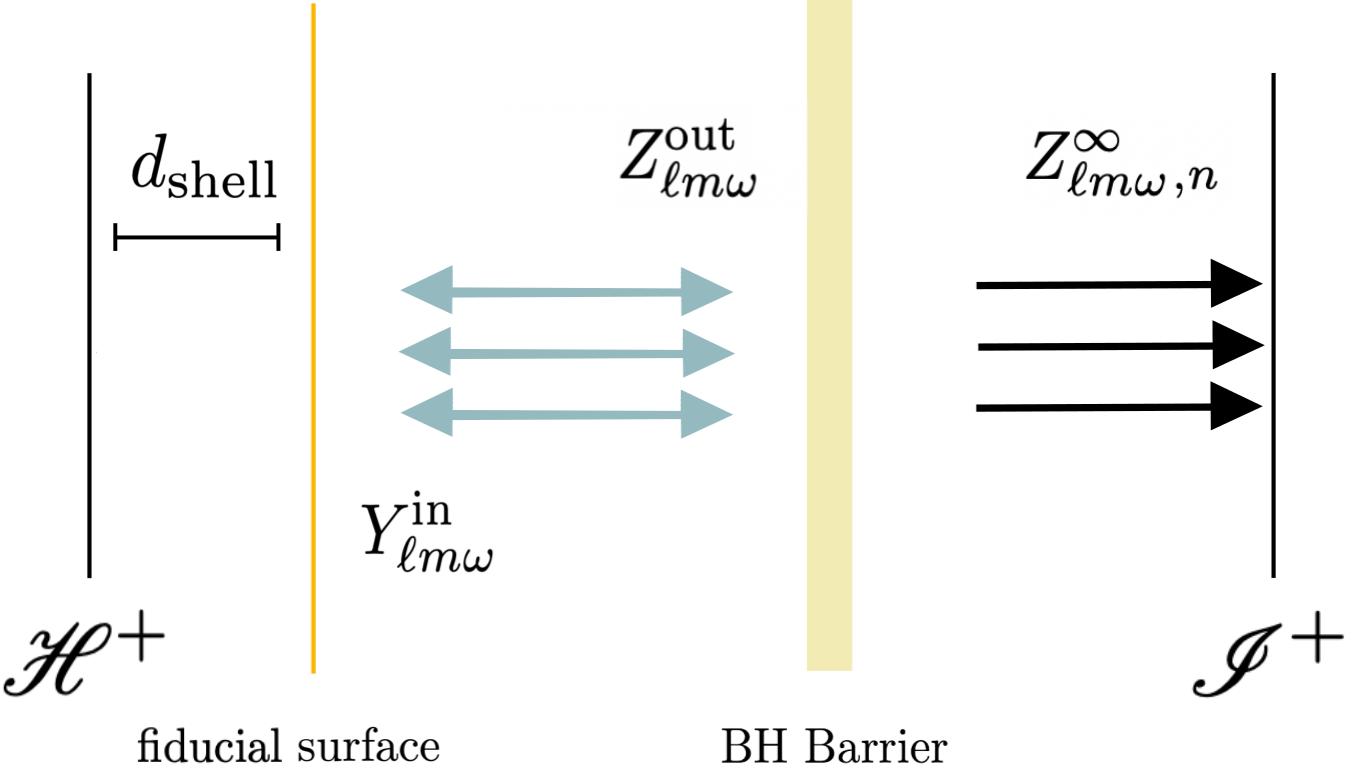}
	\caption{Sketch of the cavity formed by the BH potential barrier and the reflective shell separated by $d_\T{Shell}$ from the future BH horizon $\Hp$. The radial amplitudes of gravitational radiation (sourced by the ringdown of the perturbed BH) are denoted by $Y_{\ell m \omega},Z_{\ell m \omega}$. The GW detector is located at future null infinity $\scrip$.}
	\label{fig:Sketch_intuition}
\end{figure}

The phase information carried by the reflectivity $\RQBH$ determines the time separation between echoes. It can be deduced by studying the echo production mechanism (Fig. \ref{fig:Sketch_intuition}): Consider gravitational radiation emitted during the ringdown of the perturbed post-merger BH entering a cavity formed by the BH potential barrier and a shell of radius $r_\T{Shell}$ surrounding the BH. For each cycle that radiation undergoes within this cavity with partially reflective ``walls'', an echo leaks out traveling towards an observer at $\scrip$. Staying agnostic with regards to the shell's exact physical properties, we treat it as a fiducial surface (compare with discussions in \cite{Giddings_2016}). By simple time delay arguments, $r_\T{Shell}>r_H$, where $r_H = 2M$ is the Schwarzschild radius, to find any detectable signal reaching an outside observer in finite time. Slightly generalizing to include spinning BHs, $r_H := r_\pm = M(1\pm \sqrt{1-a^2})$, where $a$ is the dimensionless spin parameter, the time a GW takes to complete a single cycle within the cavity is given by \cite{Abedi_2017,Wang_2018}
\begin{align}
\label{equ:t_echo_0}
    \Delta t_\T{echo} = 2 r^* \big|_{r_\T{Shell}}^{r_\T{Barrier}}\,,
\end{align}
where $r^*$ is the tortoise coordinate and $r_\T{Barrier}$ marks the location of the BH potential barrier. The latter implies \cite{Wang_2018}
\begin{align}
\label{equ:t_echo_II}
    \Delta t_\T{echo} 
    \approx 2 \frac{r_+^2 +a^2M^2}{r_+-r_-}\ln{\left(\frac{M}{r_\T{Shell}-r_+}\right)} + M f(a)\,,
\end{align}
where ${r_\T{Shell}}-r_+ =: d_\T{Shell}^2\sqrt{1-a^2}/4M(1+\sqrt{1-a^2})$ and $f(a)\approx 0.335/(a^2-1) + 4.77 + 7.42(a^2-1) + 4.69(a^2-1)^2$. The free parameter in this consideration is denoted as $d_\T{Shell}$ and describes the proper distance between the fiducial shell, located at $r_\T{Shell}$, and the BH horizon at $r_H$. In the limit of $d_\T{Shell}\rightarrow 0$, one obtains an infinite time separation between the echoes. Per se, the time separation between the echo and the classical waveform is physically not constrained, i.e., echoes might appear in interferometer data without an unambiguous association to a classical merger waveform \cite{zimmerman2023rogueechoesexoticcompact}. 

Including exponential suppression and phase information, one obtains, using the convention $M=1$ and setting $a=0$,
\begin{align}
\label{equ:reflectivity_QBH}
    \RQBH =e^{-i\omega8\ln{\left[\beta(a)\right]}}
     e^{-\frac{|\omega|}{2T_{QH}}}\begin{cases} 
            1, & \omega\leq\frac{\omega_1}{2}, \\
             \left|\sin{\left( \frac{\pi\omega}{\omega_N(\alpha,a)} \right)}\right|^\delta, & \omega > \frac{\omega_1}{2},
            \end{cases}
\end{align}
where constant phase contributions have been absorbed by a new model parameter $\beta \sim d_\T{Shell}$. In the limit $\delta \rightarrow 0$ and with a suitable choice for $\beta$, Eq. \eqref{equ:reflectivity_QBH} recovers the ECO reflectivity outlined in \cite{Wang_2020, Oshita_2020}.
Throughout this work, the parameter space for the QBH reflectivity $\RQBH$ is spanned by the reflectivity parameter $\alpha, \beta, \epsilon$ and $\delta$.

As the interaction with the fiducial surface enclosing the BH has a reflective nature, one needs to additionally define a suitable boundary condition. In this work, we rely on the similarity between an ECO's surface and the fiducial shell of QBHs and postulate
\begin{align}
    \label{equ:boundary}
    \Zout \sim \gamma^{-1} \RQBH\Yin\,,
\end{align}
i.e., the ingoing radial amplitude of the GW, $\Yin$, is suppressed by a factor of $\gamma$ upon reflectivity. This notion generalizes the case of the ECO $(\gamma = 4)$ \cite{Chen_2021}. The parameter $\gamma$ is added to the reflectivity parameter space.

\section{Echo construction}
Given the reflectivity function Eq. \eqref{equ:reflectivity_QBH} and boundary condition Eq. \eqref{equ:boundary}, the echo of a GW is computed following the methodology outlined in \cite{Ma_2022}. In essence, the procedure extracts radiation heading towards the QBH horizon using the asymptotic information encapsulated in the Newman Penrose scalar extracted at future null infinity $\scrip$, $\Psi_0^\circ$. Together with the outgoing radiation in $\Psi_4^\circ$, the GW's time series measured at the detector and including the echo's strain, $\hecho_{\ell m}$, is computed. Alternatives to \cite{Ma_2022}, e.g., (analytical) approaches for templates of GW echos, have been established in \cite{Testa_2018,Maggio_2019_II}.
Physically, the generation of echoes proceeds as follows: After a binary system merges, the remnant BH is perturbed and undergoes a ringdown phase, emitting its QNM content. This process is described by the Teukolsky equation, which yields outgoing solutions towards $\scrip$ and ingoing solutions towards the future QBH horizon $\Hp$. The ingoing waves become ``trapped'' in a cavity (Fig. \ref{fig:Sketch_intuition}) formed between the QBH's fiducial reflective shell and the classical BH's potential barrier. Since neither the barrier nor the shell are perfectly reflective, a portion of the gravitational radiation leaks out during each propagation cycle in the cavity. Therefore, GWs trapped in the cavity partially fall into the BH horizon $\Hp$ and partially escape towards $\scrip$ in form of echo-like signals.

We outline relevant details of the semi-analytical echo reconstruction as well as visual intuition in \textcolor{black}{Appendix \ref{app:A}}. Further details are provided in \cite{Ma_2022}. Note that throughout this work, we compute the echo only for $\hecho_{2,\pm2}$, as the remaining harmonic strain modes are subdominant. 

\section{Detecting gravitational wave echoes}
We estimate the detectability of GW echos in data of the future space-based LISA instruments, employing the simulation and data analysis pipeline described in \cite{inchauspe2024} (for details on the calibration, see \textcolor{black}{Appendix \ref{app:B}}). We numerically determine the \gls{snr} of the echo for a set of 11 numerical relativity simulations of massive binary BH mergers (with total mass $M_\text{tot}>10^5 \Msol$) extracted from \cite{Boyle_2019}. These include \textit{SXS:BBH:1936}, \textit{0207}, \textit{0334},  \textit{1155}, \textit{1424}, \textit{1448}, \textit{1449}, \textit{1455}, \textit{1936} and \textit{2108}. The events were selected primarily based on their vanishing remnant spin. Exceptions include \textit{SXS:BBH:0334}, \textit{1155}, and \textit{2108}, which have remnant spin amplitudes $|\vec{\chi}|$ between $0.28$ and $0.68$. The latter simulations are included in the analysis to check for systematic differences in echo-related features when $a\neq0$. While the reflectivity function holds true in complete generality, the echo reconstruction in \cite{Ma_2022} is adapted to non-spinning remnants due to computational difficulties in the presence of spin. For the applications of this work, extending considerations to Kerr remnants thus introduces a small systematic \textcolor{black}{error} \footnote{Note further that non-trivial spin would induce a small line-broadening in the absorption lines corresponding the the characteristic frequencies \cite{Agullo_2021}. This effect can be captures approximately by the introduced ``sharpness''-parameter $\delta$.}. Accepting such minor corrections, our analysis is generally not limited to a specific region in parameter space and can be applied to an arbitrary waveform, given the knowledge about $\Psi_0^\circ,\Psi_4^\circ$. The extraction of the Newmann-Penrose scalars beyond $\Psi_4^\circ$ usually requires additional simulation efforts. For \textit{SXS} events, the Cauchy-Characteristic-Evolution (CCE) scheme provides the necessary tools. In this analysis, we use the CCE method \cite{Moxon_2020,moxon2021spectrecauchycharacteristicevolutionrapid} implemented in the new numerical relativity code SpECTRE \cite{Kidder_2017,spectrecode}.

Computing the SNR of the waveform echo, the reflectivity function $\RQBH$ has to be fixed. The choice of the parameters involved can significantly alter the shape of the echo in interferometer data. The phenomenological constant $\alpha$ and the cusp-parameter $\delta$ influence the location and depth of the features corresponding to the characteristic frequencies (see Fig. \ref{fig:SNR_TDI}). The temperature coefficient $T_{QH}$ (or rather $\epsilon$) and boundary suppressor $\gamma$ directly impact the amplitude of the echo. The time dilation parameter $\beta$ is irrelevant for the SNR as it only shifts the echo along the time axis, i.e., it regulates the time separation between the echo and waveform time series as well as among echoes. Generally, we assume that the time separation is large enough ($\beta$ is small enough) such that the echo does not interfere with the ringdown. 
\begin{figure}
	\centering
	\includegraphics[width=0.9\linewidth]{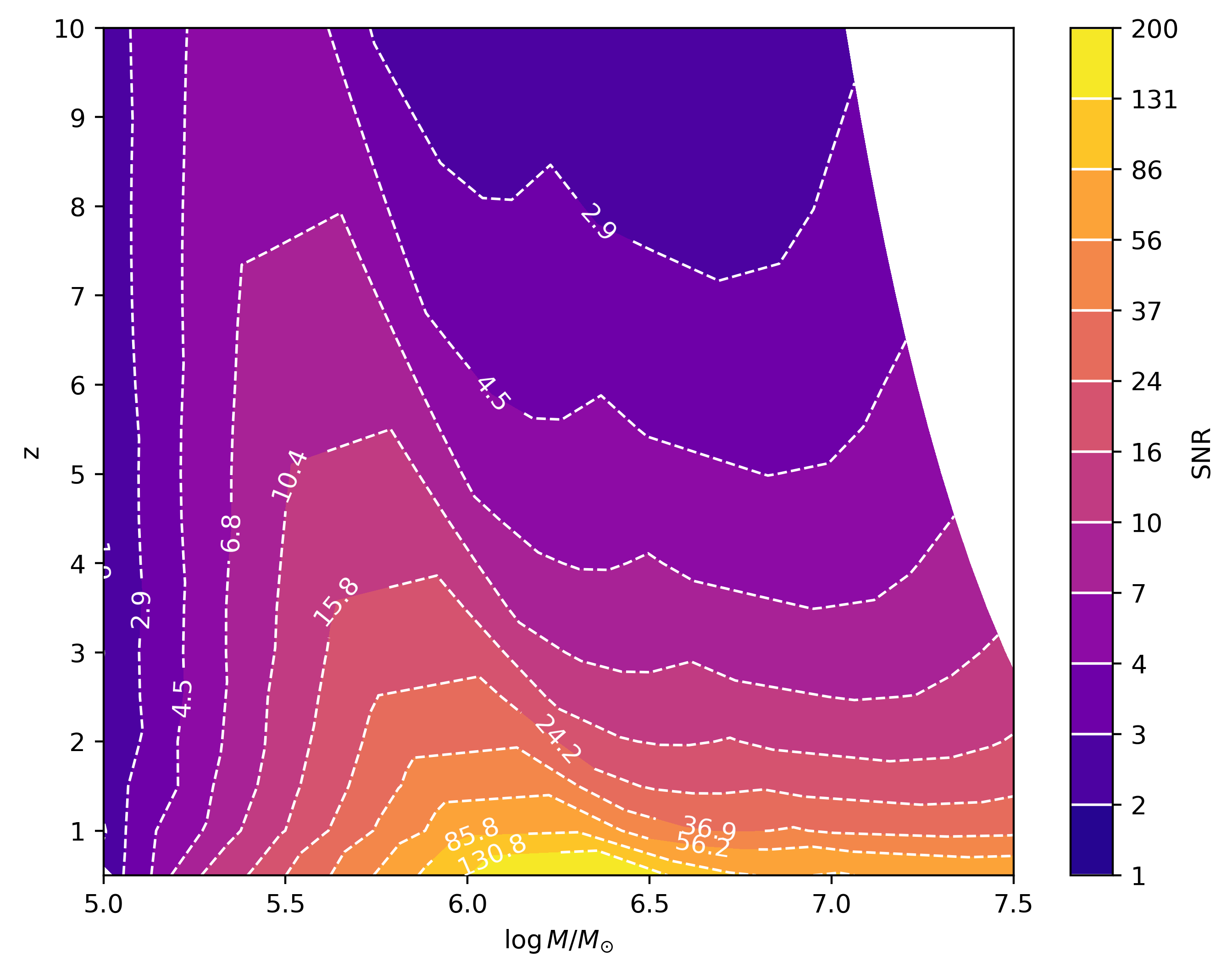}
	\caption{LISA-SNR of the echo produced by the \textit{SXS:BBH:1936} waveform over redshift and total (red-shifted) mass. We fix $ \log M_\text{tot} / \Msol = 6, z=1, \alpha = 8\pi, \delta = 0.2, \epsilon =1 ,\gamma=4$.}
	\label{fig:SNR_ECHO}
\end{figure}

We first estimate the echo SNR using the baseline parameter set $\alpha = 8\pi, \beta = 10^{-15}, \gamma = 4, \delta = 0.2, \epsilon = 1$. The echo SNR is computed as a function of mass and redshift. The result for \textit{SXS:BBH:1936} is displayed in Fig. \ref{fig:SNR_ECHO}. If the mass and redshift ($M_\text{tot} = 10^{6}M_\odot$ and $z=1$) are fixed, but $\gamma$ and $\epsilon$ are varied, the echo SNR can deviate significantly from its baseline profile. We observe in particular, that lower $\gamma$ and larger $\epsilon$ yield a larger SNR. This is to be expected as these parameters drive the overall amplitude of the echo. A visualization of the functional dependence of the SNR on $\gamma,\epsilon$ is given in \textcolor{black}{Appendix \ref{app:C}}. 

Our results indicate that, generally, there is a vast regime in the reflectivity parameter space that hints a potential echo detection with LISA. Given the expected extremely ``loud'' signals from very massive events, even very weak echoes that are strongly damped either by the boundary condition or the exponential decay of the reflectivity, exhibit high SNR $\gtrsim 10$ in the conservative baseline of \cite{inchauspe2024} for LISA. As demonstrated in \cite{inchauspe2024}, depending on the population, the number of such ultra-loud events in the regime $M_\text{tot} = 10^{6}M_\odot$ and $z=1$ can reach up to $\mathcal{O}(100)$ over an observation period of $4$ years with LISA. 
For favorable reflectivity parameters, the echo SNR can be boosted considerably. Note however that the absence of any echo detection by the LIGO collaboration so far (e.g., \cite{Westerweck_2018,uchikata2023searchinggravitationalwaveechoes}) imposes constraints on the choice of the reflectivity parameters $\gamma,\epsilon$. Due to data analysis related challenges, however, such constraints are rather vague such that for the simulated regime $\gamma\in[1,14]$ and $\epsilon\in [0.2,2.7]$ the echo amplitude remains within an unconstrained regime ($\epsilon=2,\gamma=4$ roughly corresponds to a damping factor $\gamma=0.4$ in \cite{Westerweck_2018}).

The above results are obtained independently of the time separation between the initial waveform and gravitational echo. Moreover, the cusp parameter $\delta$ and the location of the characteristic frequencies $\alpha$ play only marginal roles in the SNR computation. It is emphasized at this point though that a mere detection of the echo does not necessarily identify the signal as such, let alone unambiguously associate it with the outlined QBH phenomenology. However, as we will now demonstrate, upon echo detection, further analysis of the captured signal might be more conclusive in this regard. 

\section{Measuring Characteristic Frequencies - a Smoking Gun for Black Hole Physics}
\begin{figure}
	\centering
	\includegraphics[width=0.9\linewidth]{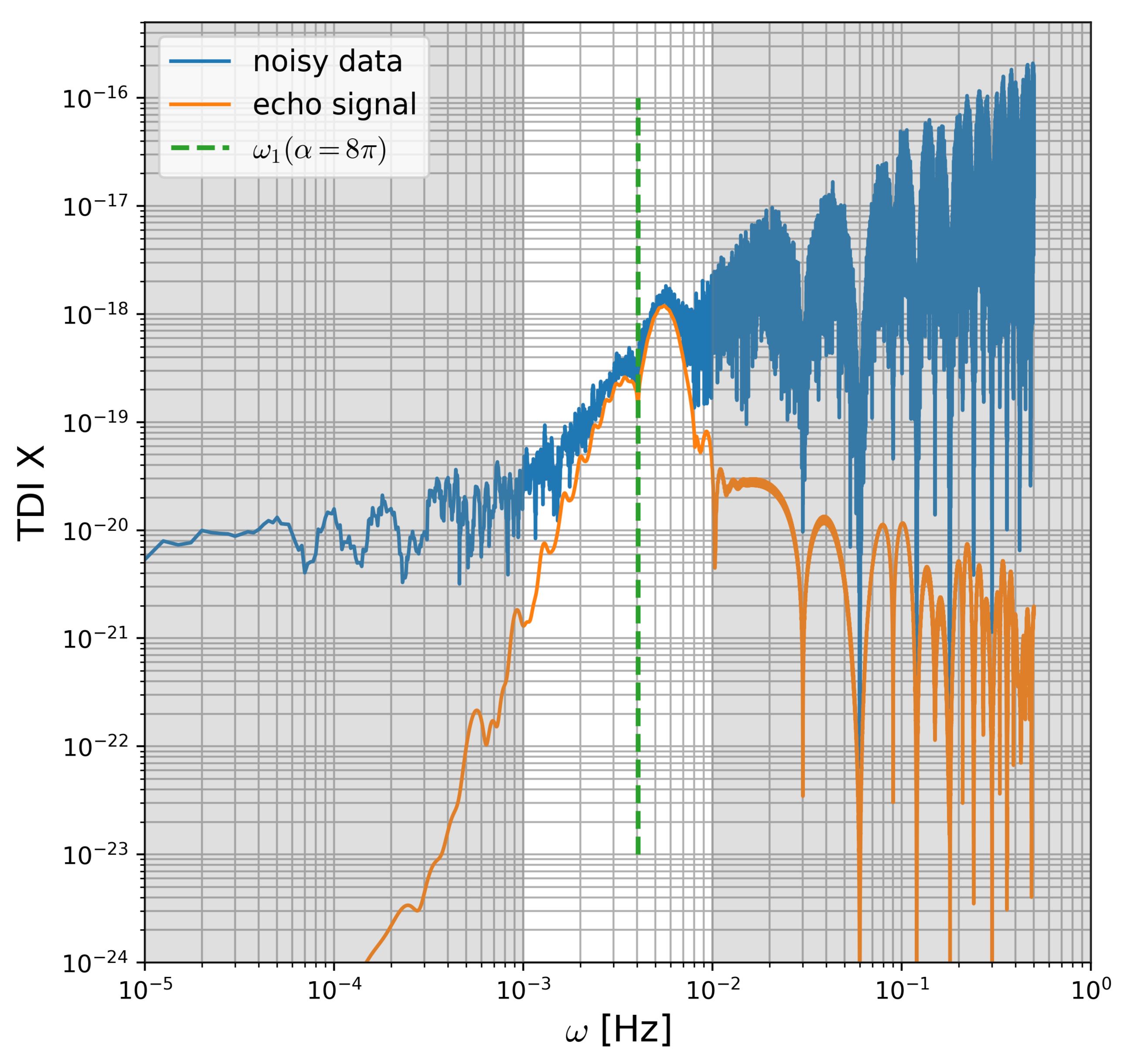}
	\caption{LISA TDI X channel data for the echo of \textit{SXS:BBH:1936} (including noise) with the baseline parameters for the reflectivity and $\delta = 0.5$ at redshift $z=1$ and $\log M_\text{tot}/M_\odot = 6$, including noise. The orange graph represents the signal without noise. The dashed line indicates the location of the first characteristic frequency of the QBH for the given mass and $\alpha$. The gray-shaded frequency domains are excluded from the fitting procedure. }
	\label{fig:SNR_TDI}
\end{figure}

Assuming the detection of an echo with LISA, crucial information can be extracted from its unique features in the interferometer data. In this analysis, we focus on features within the data collected in the \gls{tdi} channels and associated with the characteristic frequency $\omega_N$. Detecting characteristic frequencies constitutes a smoking gun for BH physics as their measurement would provide a direct probe of the area quantization of the BH event horizon. These frequencies correspond to roots of the echo's strain in frequency space, which manifest as cusps (due to $\delta >0$) in the TDI data, as displayed in Fig. \ref{fig:SNR_TDI}. 

To obtain a first estimate for the detectability of the characteristic frequency, determined by the fundamental constant $\alpha$ (and spin $a$), with LISA, we compute the uncertainty of the associated feature within the TDI X signal. Concretely, we fit the echo's transfer function, 
\begin{align}
\label{equ:transfer_main}
    \mathcal{K}(\omega)=\frac{- \RQBH/(\gamma \Dout)}{1-\RQBH \RBH} \,, &&\text{with} && \hecho_{\ell m}(\omega)\sim\frac{1}{\omega^2} \mathcal{K}(\omega)\,,
\end{align} 
to the simulated TDI X data in frequency space using a non-linear weighted least-squares and a Markov-Chain-Monte-Carlo (MCMC) scheme. Here, $\RQBH,\RBH$ are the reflectivities of the BH and its potential barrier while $1/\Dout$ denotes the transitivity of the latter. A sketch of the derivation of \eqref{equ:transfer_main} is provided \textcolor{black}{in Appendix \ref{app:A}}. The TDI data used for fitting includes simulated LISA noise following a conservative noise model. The same model has been used in \cite{inchauspe2024}. The fit function is informed about the noise's statistics via its Power Spectral Density (PSD). Fig. \ref{fig:SNR_TDI} displays an instance of noisy data (including the TDI X features of an echo) and the pure echo signal without the corresponding merger waveform.
The uncertainty provided by the fitting schemes is subsequently denoted by $\sigma_\T{fit}$, representing the standard deviation error of the fit parameter $\omega_1$.

The fit is performed with respect to $\omega_1$ and $\delta$. The former can be converted to $\alpha$ using Eq. \eqref{equ:characteristic_frequency}.
Except for $\epsilon, \gamma$, the remaining reflectivity parameters are irrelevant to the fitting procedure as they do not modulate the signal's amplitude. For our analysis, we chose $\epsilon=1$ as it represents the phenomenologically favored value (where $T_\T{QH}=T_\T{H}$) given the motivation above. For $\gamma$, we recover a boundary condition similar to those of ECOs by choosing $\gamma = 4$. Due to their direct impact on the echo's amplitude, larger $\epsilon$ and smaller $\gamma$ generally yield smaller uncertainties for the identification of $\omega_1$. The latter holds true as long as the fitting precision is limited by noise, i.e., roughly for $\epsilon\lesssim1.5$ and $\gamma\gtrsim5$. For $\epsilon>1.5$ and $\gamma<5$, the echo's signal dominates the conservative noise model such that the fidelity of the estimate of $\omega_1$ is constraint by the frequency resolution of the TDI channel only. 

The uncertainties are computed for $\alpha \in [4 \log 2,8\pi]$ and $\delta \in [0.05,0.6]$. Our results indicate that for the majority of the outlined parameter space, the echo's signal amplitude in the TDI X channel is large enough to resolve the characteristic frequency within $5\sigma_\T{fit}$ of the true value. In the tested regime (fixing $M/M_\odot = 10^6$ and $z=1$), the characteristic frequencies lay between $1$ and $4$ mHz. Given a frequency resolution $\Delta\omega_\T{TDI}$ of roughly $2.5$ $\mu$Hz, we obtain uncertainties between 5 and 100 $\Delta\omega_\T{TDI}$. In particular, we find that for large $\alpha$'s (around $8\pi$), or equivalently $\omega_1 \approx 4$ mHz, the uncertainties are small throughout the tested values of $\delta$. In the parameter regime where the frequency resolution acts as the limiting factor, uncertainties (i.e., $\sigma_\text{fit}$) could in principle be reduced by further refining the frequency resolution of the corresponding TDI channel, $\Delta \omega_\T{TDI}$. Note, however, that the TDI frequency resolution is constrained by the duration of the signal, i.e., the duration of the echo.

We further find that there is a trade off with respect to the reflectivity parameter when the uncertainty is computed: Smaller $\alpha$'s shift the characteristic frequencies in the TDI data towards the low-frequency noise-dominated regime, resulting in larger uncertainties unless the features are more pronounced, i.e., $\delta$ is larger. The presence of higher-order $\omega_N$ can only mildly compensate the noise domination of the feature associated with $\omega_1$ as there is no way to distinguish $\omega_2$ and $\omega_1$ if the latter is hidden behind the noise. Therefore, it is to be expected that the small-$\alpha$ sector of the QG theory space is generally less tightly constraint by echo searches. 

The above results hold true for all tested waveforms listed above. For \textit{SXS:BBH:1936}, they are displayed in \textcolor{black}{Appendix \ref{app:D}}, together with further details on the fitting scheme applied here. 

\section{Discussion}
\label{sec:discussion}
We demonstrated that, even under a conservative baseline and moderate echo amplitude, LISA will be capable of detecting gravitational wave echoes originating from non-trivial QBH reflectivity. Notably, the selected range of reflectivity parameters also encompasses values commonly used to describe ECOs, allowing for the extension of our results to both scenarios. Consequently, a non-detection of the echo would significantly constrain both BH and ECO phenomenology. The forecast regarding the echo's detectability is independent of the exact QNM content and has been analyzed across 11 numerical relativity simulations. For the tested mergers with non-negligible remnant spin, we did not observe any indications of relevant modifications regarding the echo when compared to the spin-less remnants. 

Based on this detection prospect, we showed that across a wide range of masses, redshifts, and binary BH mergers, the features in the LISA TDI data corresponding to characteristic frequencies are likely to be detectable to high precision. Given this capabilities, LISA may be able to discriminate between the outlined echo phenomenologies and, even more significantly, unambigously identify the echo as such, given the uniqueness of the characteristic frequency absorption line. This result implies that future GW instruments like LISA establish unprecedented probes of spacetime's quantum structure magnified by BHs. A direct measurement of the characteristic absorption frequency of BHs would further provide unique test of the BH mass-area relation and offer stringent constraints on QG theories. Non-detection of absorption frequencies in echo data implies a more membrane-inclined QBH phenomenology, similar to ECOs. Generally speaking, detecting echo signals provides crucial insights into BH physics and potentially offers direct evidence for QG effects, making it imperative to incorporate the search for GW echoes into the scientific agenda of future GW experiments. 

We acknowledge that several key assumptions in this work -- particularly the validity of the chosen boundary condition as well as the reflectivity function $\RQBH$ -- require further refinement backed by experimental and theoretical arguments. However, for the purposes of this investigation, these assumptions are sufficient, and a more rigorous derivation of the mechanisms behind $\RQBH$, including for instance quantum information theoretical considerations, is left for future work.
Finally, although our analysis has proven robust against substantial variations in certain model parameters, we address the issue of model dependence in echo searches in a companion paper \cite{Maibach2024}.



\section*{Acknowledgements}
The authors wish to thank  Do\u{g}a Veske for productive discussion and thank the LISA Simulation Working Group and the LISA Simulation Expert Group for the lively discussions on all simulation-related activities.
LH would like to acknowledge financial support from the European Research Council (ERC) under the European Unions Horizon 2020 research and innovation programme grant agreement No 801781. LH further acknowledges support from the Deutsche Forschungsgemeinschaft (DFG, German Research Foundation) under Germany's Excellence Strategy EXC 2181/1 - 390900948 (the Heidelberg STRUCTURES Excellence Cluster). The authors thank the Heidelberg STRUCTURES Excellence Cluster for financial support and acknowledge support by the state of Baden-W\"urttemberg, Germany, through bwHPC. 
Research at Perimeter Institute is supported in part by the Government of Canada through the Department of Innovation, Science and Economic Development and by the Province of Ontario through the Ministry of Colleges and Universities. This material is based upon work supported by the National Science Foundation under Grants No. PHY-2407742, No. PHY- 2207342, and No. OAC-2209655 at Cornell. Any opinions, findings, and conclusions or recommendations expressed in this material are those of the author(s) and do not necessarily reflect the views of the National Science Foundation. This work was supported by the Sherman Fairchild Foundation at Cornell. This work was supported in part by the Sherman Fairchild Foundation and by NSF Grants No. PHY-2309211, No. PHY-2309231, and No. OAC-2209656 at Caltech.

\appendix

\section{Echo reconstruction and transfer functions}
\label{app:A}
We compute the echo of a given GW following the methodology outlined in \cite{Ma_2022}, using the Newman Penrose scalars $\Psi_0^\circ$ and $\Psi_4^\circ$ at future null infinity $\scrip$. We thereby distinguish between the initial (classical) waveform, computed via the asymptotic scalar $\Psi_4^\circ$, and the radiative content sourcing the echoes, which is encapsulated in $\Psi_0^\circ$. Both quantities can be computed at $\scrip$ using numerical relativity codes. To connect this asymptotic information to the ingoing solution of the Teukolsky equation, the \textit{hybrid approach} is utilized (see \cite{Ma_2022} and references therein): First, the corresponding GW is computed at the full spacetime boundary ($\scrip \cup \scrim$) imposing a no-ingoing boundary condition, the so-called \textit{up-solution}. In this case, solving the Teukolsky equation, the asymptotic behavior of the radial coefficients of $\Psi_4$ and $\Psi_0$, respectively, is described by 
\begin{subequations}
\label{equ:BH_teukolsky}
    \begin{align}
       \,_{-2}R_{\ell m \omega} 
       \sim\begin{cases} 
            r^3 Z^\infty_{\ell m \omega}e^{i\omega r^*}, & r^* \rightarrow +\infty, \\
             Z^{\T{out}}_{\ell m \omega} e^{i\omega r^*} + \Delta^2 Z^{\T{in}}_{\ell m \omega}e^{-i\omega r^*}, & r^*\rightarrow - \infty,
            \end{cases}\\
        \,_{+2}R_{\ell m \omega} 
       \sim\begin{cases} 
            r^{-5}Y^\infty_{\ell m \omega} e^{i\omega r^*}, & r^* \rightarrow +\infty, \\
             Y^{\T{out}}_{\ell m \omega} e^{i\omega r^*} + \Delta^{-2} Y^{\T{in}}_{\ell m \omega}e^{-i\omega r^*}, & r^*\rightarrow - \infty,
            \end{cases}
    \end{align}
\end{subequations}
where $r^*\rightarrow - \infty$ describes the limit towards the future BH horizon, $\Hp$, and $r^*\rightarrow \infty$ towards null infinity $\scrip$. The asymptotic information for $\Psi_4^\circ$ ($\Psi_0^\circ$) is encoded in $\Zinf$ ($\Yinf$) while $\Zin,\Zout$ ($\Yin, \Yout$) denote the amplitudes traveling towards and away from the horizon $\Hp$ but within the cavity described in the \textcolor{black}{main text, Fig. \ref{fig:Sketch_intuition}}. The coefficients $Z_{\ell m \omega}, Y_{\ell m \omega}$ are related via the Teukolsky-Starobinsky (TS) relations \cite{Teukolsky_1974, Starobinsky:1973aij}
\begin{align}
    \frac{4\omega^4}{\overline{C}}\Yinf&=\Zinf\,,\notag\\
    \Yin&=\frac{D}{C}\Zin \,,
\end{align}
where
\begin{align}
    C&=(\ell-1)\ell(\ell+1)(\ell+2)+12i\omega\notag\,,\\
    D&=64i\omega(128\omega^2+8)(1-2i\omega)\,,    
\end{align}
and the reflectivity coefficients of the BH potential barrier, that are $1/\Dout,\Din/\Dout$ for outgoing and $1/\Cout, \Cin/\Cout$ for ingoing radiation, see \cite{Ma_2022, Hughes_2000, Xin_2021}. The latter are well-defined for multiple BH solutions (including the here considered case of linearly perturbed Schwarzschild BHs) and can be numerically determined using the Black-Hole Perturbation Toolkit \cite{BHPToolkit}.

Using the radial solutions \eqref{equ:BH_teukolsky} together with the reflectivity coefficients for QBH computed \textcolor{black}{in the main text}, the radiation bouncing of the reflective shell, $\Zout$, can be computed using the ingoing contribution (with respect to the BH horizon) to $\Psi_0$ in the limit $r^*\rightarrow-\infty$, i.e., $\Yin$ \cite{Ma_2022}. Staying agnostic with respect to the boundary conditions for the reflectivity of $\Yin$ at the QBH's fiducial shell, we assume, inspired by \cite{Chen_2021},
\begin{align}
\label{equ:boundary_equ}
    \Zouteco = \frac{(-1)^{\ell + m +1 }}{\gamma} \RQBH \Yineco \,,
\end{align}
thereby adding another reflectivity parameter $\gamma$ to our QBH model. The coefficient $\Yineco$ can be extracted from $\Yin$ by isolating contributions associated solely with the ringdown phase of the waveform. These sections are computed via the fitting procedure outlined in \cite{Ma_2022}, where a minimal mismatch between the numerical relativity waveform for $\Yin$ and an analytical expression for the ringdown modes,
\begin{align}
\label{equ:overtones}
    Y^\text{in}_{22}(v>v_\Sigma) = \sum_{n=0}^{n_\text{max}}\left(\mathcal{A}_ne^{-i\omega_nv}+\mathcal{B}_ne^{i\overline\omega_nv}\right)\,,
\end{align}
is computed with the time marking the onset of the ringdown, $v_\Sigma$, as a free parameter. The extracted time parameter then determines the Planck Filter $\mathcal{F}(v)$,
\begin{align}\label{equ:taper}
  \mathcal{F}(v,\Delta v, v_\Sigma) =  
  \begin{dcases*} 
  0, & $v < v_\Sigma - \Delta v$ \\ 
  \left(\exp{\chi+1}\right)^{-1} &  $v_\Sigma -\Delta v <v <v_\Sigma$ \\ 
  1, & $v>v_\Sigma $
  \end{dcases*} \,,
\end{align}
such that the ringdown information in $\Yin$ is determined via
\begin{align}
\label{equ:filter_Y}
    Y^\T{in QBH}_{\ell m} (v) = Y^\T{in}_{\ell m} (v) \mathcal{F}(v) + \T{Const.} \cdot \big(1-\mathcal{F}(v)\big)\,.
\end{align}
The function $\chi$ is given by $\chi(v,\Delta v, v_\Sigma) = \left(\frac{\Delta v}{v-v_\Sigma} + \frac{\Delta v}{v-v_\Sigma + \Delta v}\right)$. The Planck-taper window function \eqref{equ:taper}  is applied to reduce spectral leakage associated with the abrupt onset of the ringdown phase of the binary merger.

\begin{figure}
	\centering
	\includegraphics[width=0.99\linewidth]{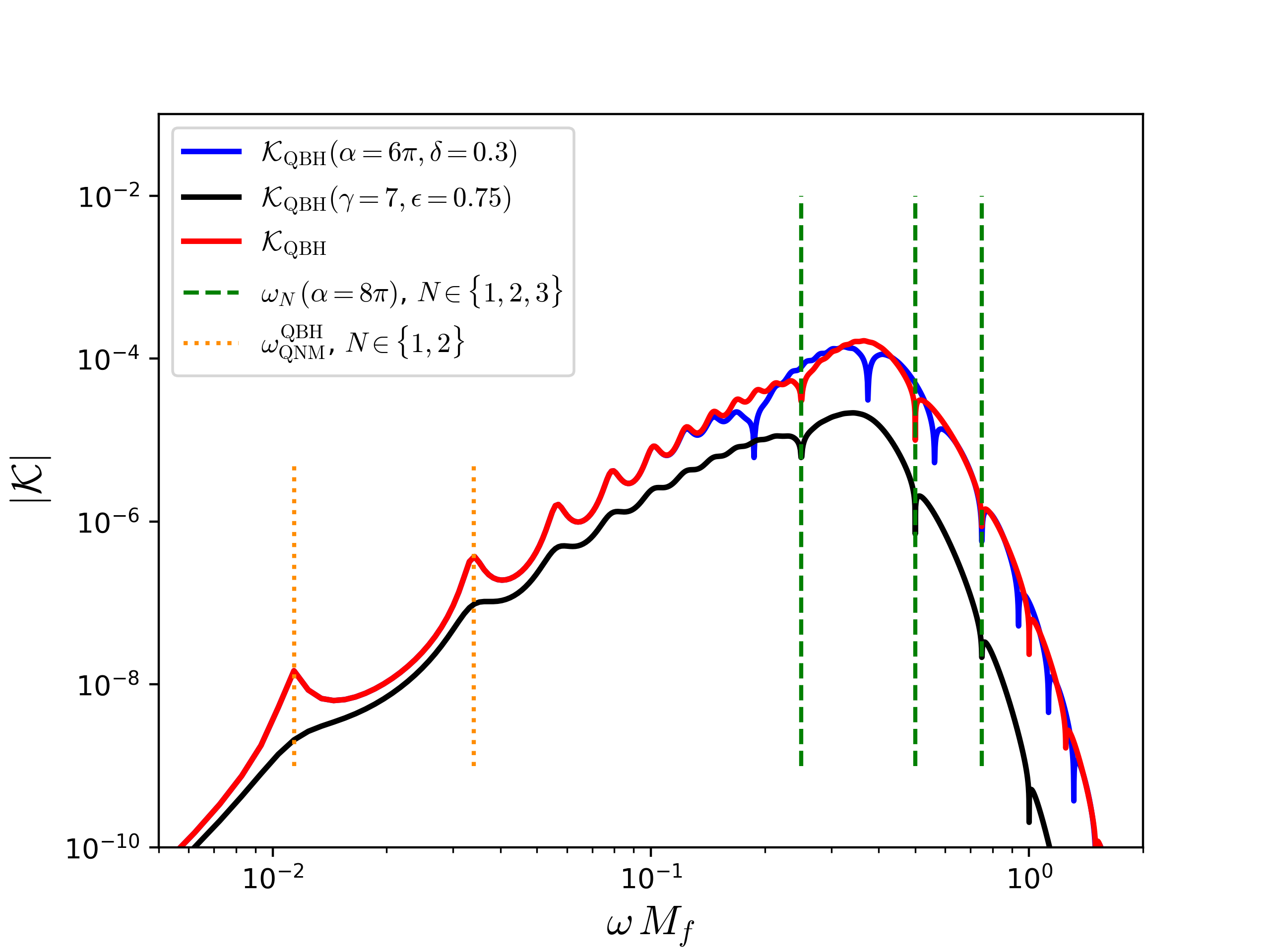}
	\caption{Transfer function for the baseline reflectivity parameter $\alpha = 8 \pi,\beta = 10^{-15}, \gamma = 4, \delta = 0.2, \epsilon=1$. Any variations in these parameters are indicated.}
	\label{fig:TRANSFER}
\end{figure}

Eq. \eqref{equ:boundary_equ} describes the radiation reflected by the horizon traveling towards the BH potential barrier. Picking up a factor of $1/\Dout$ upon transmission through the barrier, the resulting expression can be converted into gravitational strain by dividing by $\omega^2$. Decomposed into modes, one obtains
\begin{align}
\label{equ:hecho}
    \hecho &= \sum_n \sum_{\ell, m} \,_{-2}Y_{\ell m}(\theta, \phi) \hecho_{\ell m, n}(u)
\end{align} 
where the Fourier transform of the modes $\hecho_{\ell m, n}(u)$ is given by
\begin{align}
\label{equ:final_echo}
\hecho_{\ell m, n}(\omega)=&\frac{1}{\omega^2}\frac{C}{D \Din}\left(\RQBH\RBH\right)^n \notag\\&\cdot\mathfrak F\left\{C^\T{in}_{\ell m} \Psi_{0, \ell m}(v)\mathcal{F}(v) \right\}\,.
\end{align}
Here, we define $\mathfrak F (\Psi_{0,\ell m}) =: \Yinf$ and $\mathfrak F (\cdot)$ denotes the Fourier transform. The index $n$ denotes the number of echoes included in the strain time series \eqref{equ:hecho}. The coefficient $\RBH$ denotes the reflectivity of the BH barrier defined as in \cite{Ma_2022}. It is instructive to introduce a transfer function and to rewrite Eq. \eqref{equ:final_echo} as
\begin{align}
    \label{equ:echo_with_transfer}
    \hecho_{\ell m}(\omega)=&\frac{1}{\omega^2} \mathcal{K}(\omega)\Yineco 
\end{align}
with 
\begin{align}
\label{equ:transfer}
    \mathcal{K}(\omega) =& \sum_n \frac{C}{D \Din}\left(\RQBH\RBH\right)^n\notag\\
    =& \frac{(-1)^{\ell + m +1 } \RQBH}{1-\RQBH \RBH} \frac{1}{\gamma \Dout}\,.
\end{align}
The transfer function is fundamentally frequency dependent and contains poles at $1-\RQBH(\omega) \RBH(\omega)=0$. The frequencies for which the latter is satisfied describe the QNMs of the QBH. 
An example of the transfer function is depicted in Fig. \ref{fig:TRANSFER}. In the latter, we mark the QBH's QNMs as well as the characteristic BH frequencies $\omega_N$ dictated by $\RQBH$ in the numerator of \eqref{equ:transfer}, appearing as zeros. Note that the poles in Fig. \ref{fig:TRANSFER} do not correspond exactly to the QNMs but to the resonances of the system described by Fig. \ref{fig:Sketch_intuition} of \textcolor{black}{the main text}. The QNMs reside very close to these resonances in frequency space.

\begin{figure}
	\centering
	\includegraphics[width=0.90\linewidth]{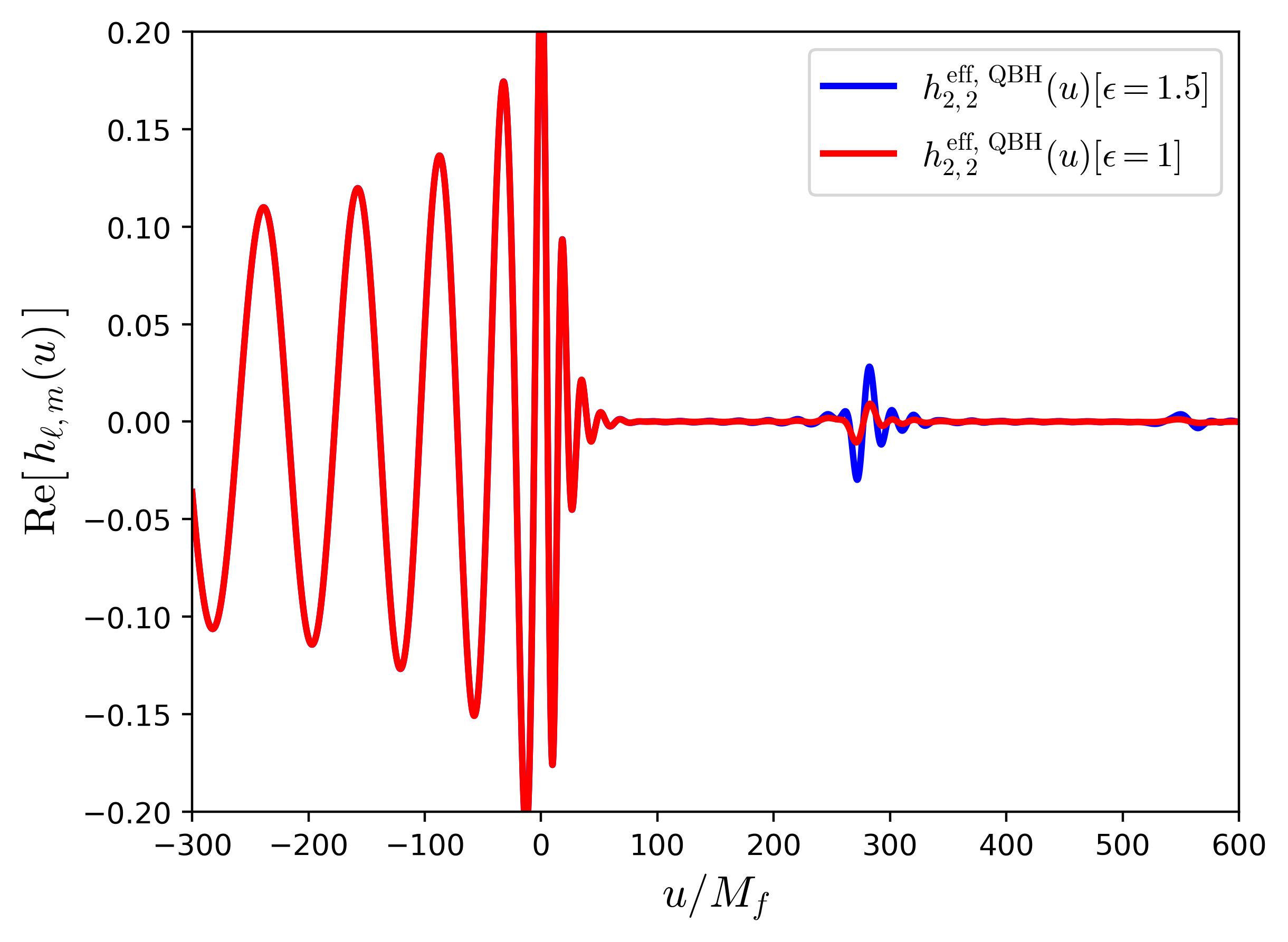}
	\caption{Exemplary waveform with echo computed for \textit{SXS:BBH:1936} with the baseline reflectivity parameter $\alpha = 8 \pi,\beta = 10^{-15}, \gamma = 4, \delta = 0.2$. The exponential suppression parameter $\epsilon$ is chosen as indicated.}
	\label{fig:ECHO}
\end{figure}

The echoes resulting from this procedure are computed numerically. The strain time series given by Eq. \eqref{equ:hecho} are added to the classical waveform. The result is exemplarily sketched in Fig. \ref{fig:ECHO} for event \textit{SXS:BBH:1936} and a random selection of reflectivity parameters. 

\section{Waveform SNR}
\label{app:B}
\begin{figure}
	\centering
	\includegraphics[width=1\linewidth]{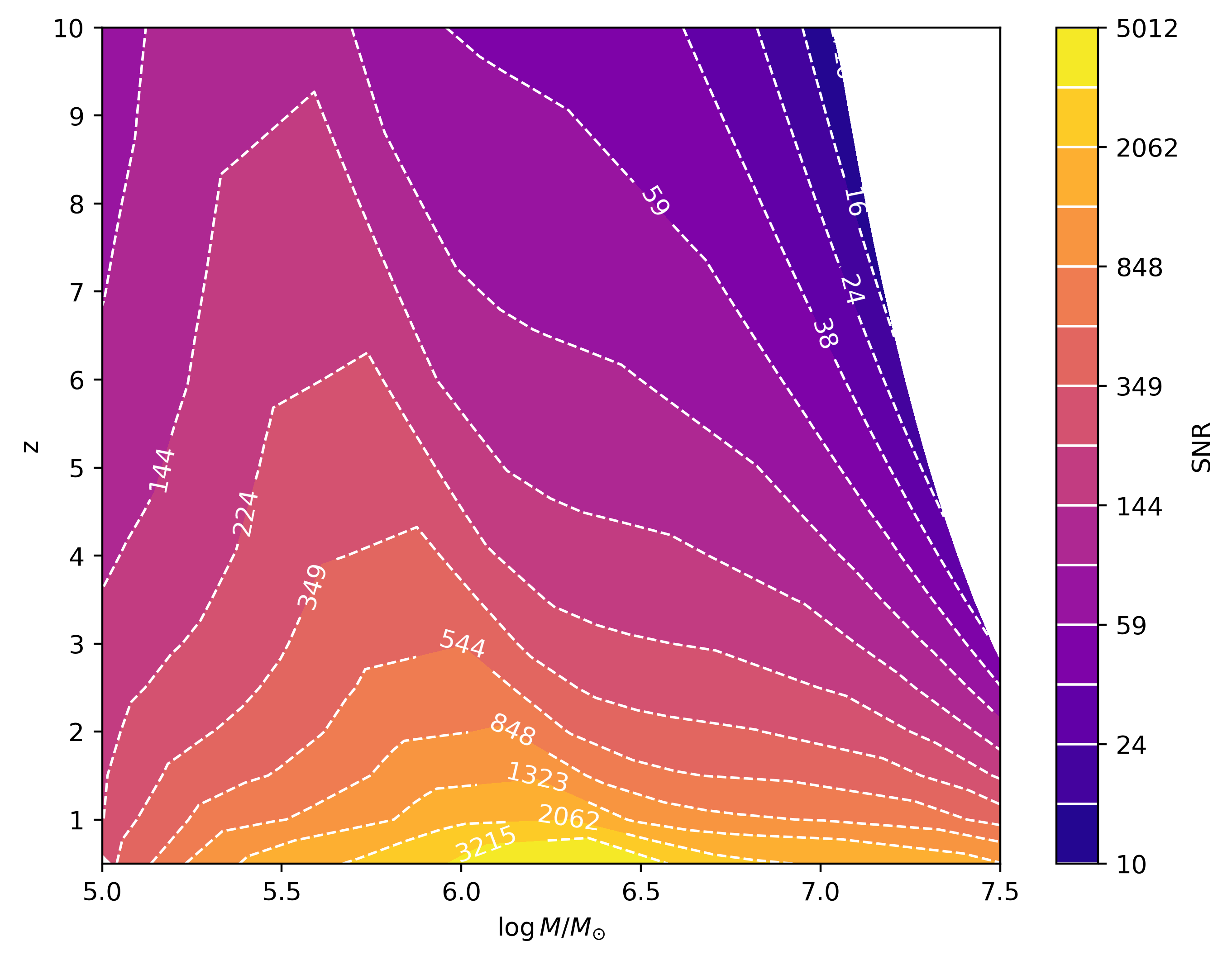}
	\caption{LISA-SNR of the \textit{SXS:BBH:1936} waveform for different redshifts and total (redshifted) masses. The SNR is computed following \cite{inchauspe2024}.}
	\label{fig:SNR}
\end{figure}

Throughout the analysis, we investigate the signals captured by the TDI X channel of the LISA instrument. We emphasize, that there is no particular preference of the X channel. Alternatively, other channels with sufficient signal SNR as well as the full strain data would be eligible for the detectibility estimation depending on the orientation with respect to the GW source. Regarding the sky position and other orientation-dependent features relevant to the SNR, we adapt the conservative baseline as in Table I of \cite{inchauspe2024}. For reference, the SNR of the full waveform of event \textit{SXS:BBH:1936} picked up by LISA is displayed in Fig. \ref{fig:SNR}, as a function of redshift and redshifted mass. Note that we do not reproduce the plot of the conservative baseline of \cite{inchauspe2024}, as different simulation parameters (i.e., binary spins, mass ratio, etc.) have been chosen in there. 




\section{Echo SNR}
\label{app:C}
\begin{figure}
	\centering
	\includegraphics[width=1\linewidth]{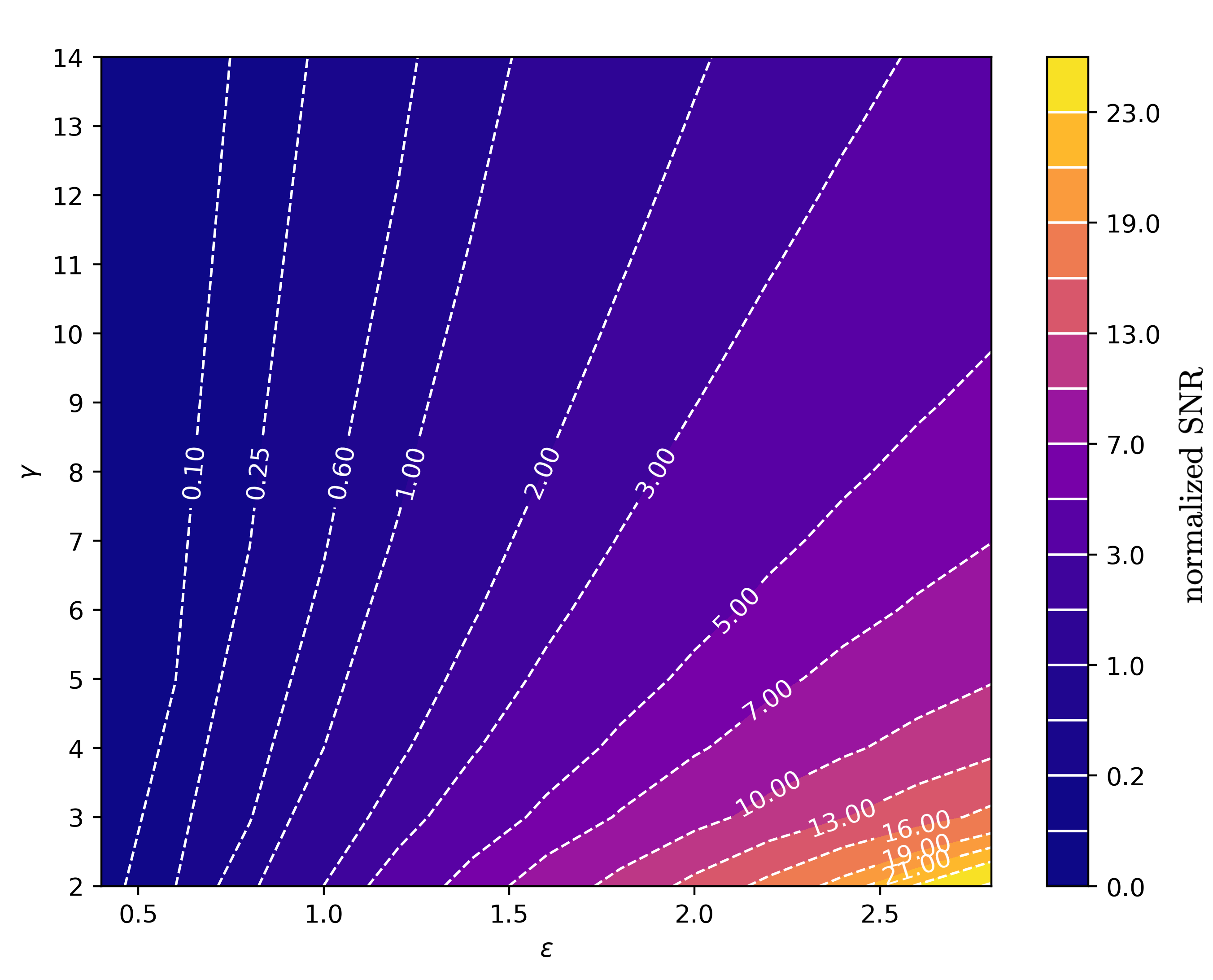}
	\caption{Normalized SNR for the echo of \textit{SXS:BBH:1936}. We fix $ \log M / \Msol = 6, z=1, \alpha = 8\pi, \delta = 0.2$ and vary $\epsilon,\gamma$. The resulting SNR is normalized to the SNR resulting from the choice $\epsilon=1,\gamma=4$, which is $\approx 86$.}
	\label{fig:ECHO_PARAM}
\end{figure}

We exemplarily display the echo SNR for the numerical relativity simulation\textit{SXS:BBH:1936} for fixed mass and redshift in Fig. \ref{fig:ECHO_PARAM}. Thereby, we normalize the SNR with respect to the baseline reflectivity parameters ($\gamma=4, \epsilon=1$). For a given merger determined by redshift and mass, the parameter-dependent echo SNR is given by the product of the SNR value in Fig. \ref{fig:SNR_ECHO} of the main text and the corresponding factor in Fig. \ref{fig:ECHO_PARAM}, depending on the choice of reflectivity. For instance, for a merger at redshifted mass $\log M / \Msol = 6$ and redshift $z=1$, the SNR for $\epsilon =1$ and $\gamma =6$ is given by $\T{SNR} \approx 86 \cdot 0.6 \approx 52$.

\section{Fitting scheme for characteristic frequencies}
\label{app:D}

To extract the characteristic frequencies from the echoes' frequency space data, an MCMC fitting scheme is applied, matching the Fourier transform of the echo stain time series with the noise-contaminated data simulated for a realistic LISA measurement. The shape of the echo strain in frequency space is captured with sufficient accuracy through the transfer function \eqref{equ:transfer}. Due to the numerical simplicity of computing the transfer function, we chose a rescaled version of the latter as the fitting function in our procedure. Thereby, the rescaling parameter constitutes an additional fit parameter.

\begin{figure}
	\centering
	\includegraphics[width=1\linewidth]{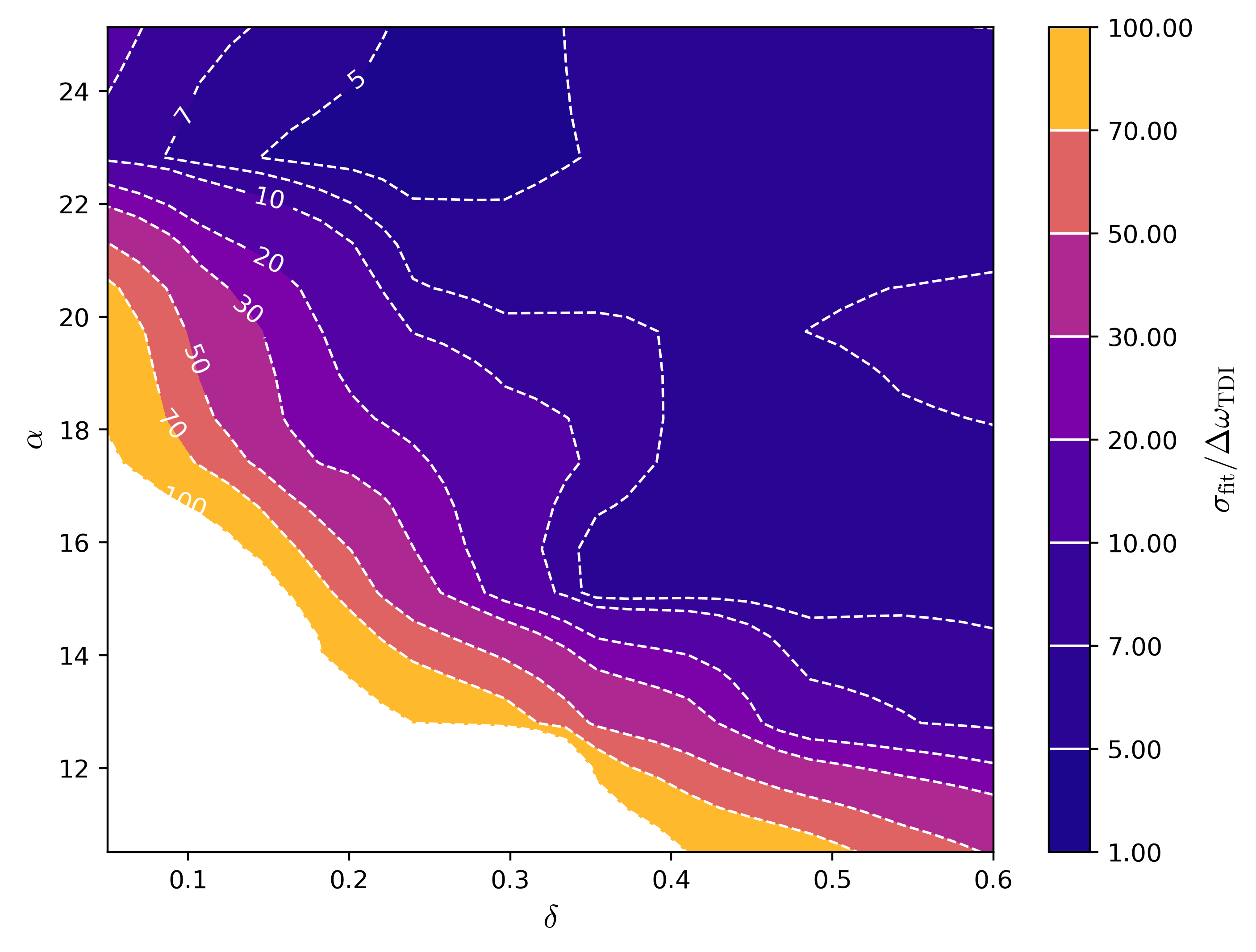}
	\caption{Uncertainty of the characteristic frequency normalized by the TDI frequency resolution ($\approx 2.5$ $\mu$Hz) and extracted from a fit to the TDI data of echo of the simulated waveform \textit{SXS:BBH:1936}. We fix $ M/M_\odot = 10^6, z=1, \epsilon=1, \gamma=4$ and vary $\alpha,\delta$.}
	\label{fig:SNR_PEAK_PARAMS}
\end{figure}

To stress the robustness of our analysis, we perform the analysis over 20 different realizations of the noise. We find that, empirically, the scattering of the 20 recovered frequencies $\omega_1$ are statistically consistent with the theoretical errors $\sigma_\T{fit}$. Further, to provide reliable estimates for the detectability, we compute the uncertainty $\sigma_\T{fit}$ via a weighted least squares method and a deeper Bayesian MCMC analysis. Thereby, both MCMC and fitting scheme are executed without prior knowledge about the characteristic frequencies' location. The MCMC is initiated based on a no-cusp-scenario, i.e., $\alpha=0=\delta$. The bounds for $\alpha$ and $\delta$ for both methods are set to extend over $\alpha \in [4\log 2, 8\pi]$ and $\delta\in (0.05, 0.6]$. The tested parameter range of $\alpha$ is chosen to include all phenomenological constants appearing in literature. For $\delta$, we chose a sufficiently large region covering a wide spread of phenomenologies. \textcolor{black}{The results of the simulation are displayed in Fig. \ref{fig:SNR_PEAK_PARAMS}}. The uncertainties obtained from the MCMC are consistent with an expected likelihood approach in which the likelihood function of the MCMC is weighted by the noise PSD but the noise realization within the data converges to zero.

It is further noted that we choose an event with $\log M/M_\odot = 6$ and $z=1$ in our analysis to guarantee a sufficient echo SNR. This choice fixes the peak of the echo's frequency signal in the frequency domain (compare \textcolor{black}{Fig. \ref{fig:SNR_TDI}}). Generally, the remnant's mass can have a considerable impact on the uncertainty. Increasing (Decreasing) the mass shifts the characteristic frequency to the lower (higher) frequencies as $\omega_N \sim 1/M$. For roughly $\log M/M_\odot \gtrsim 7.5$ ($\log M/M_\odot \lesssim 5.5$), it exits the sensitivity band of LISA. Thus, coincidentally, the mass domain for which echos are strong and for which the characteristic frequency can potentially be identified aligns.

\twocolumngrid

\printglossary[type=\acronymtype]
\bibliography{ref}

\end{document}